\documentclass[a4paper]{PoS}
\pdfoutput=1
\usepackage{amssymb}
\usepackage{graphicx}
\usepackage{amsmath}
\usepackage{mathrsfs}
\usepackage{fontenc}
\usepackage{times}
\usepackage{mathptmx}

\title{On the gravitational quantum states of helium atoms in the gravitational field of a cold neutron star}

\ShortTitle{On the gravitational quantum states of helium atoms}

\author{\speaker{O.D. Dalkarov}\\
P.N. Lebedev Physical Institute, 53 Leninsky
prospect, 117924 Moscow, Russia.\\
E-mail: \email{dalkarov@sci.lebedev.ru}}
\author{E.A. Kupriyanova\\
P.N. Lebedev Physical Institute, 53 Leninsky
prospect, 117924 Moscow, Russia.\\
E-mail: \email{kupr\_i\_k@mail.ru}}

\abstract{A study of gravitational properties of matter presents a fundamental interest. The possibility of investigation of quantum gravitational states of matter by the example of helium atom is shown.

The capability of the existence of helium quantum states in the gravitational field of a cold neutron star is examined. Observation of such states is done with the help of rotating neutron star's magnetic field. Periodically changing magnetic field induces transitions between gravitational states of helium atom and leads to the appearance of gravitational transitions' spectral lines in gigahertz frequency range.}

\FullConference{The 34th International Cosmic Ray Conference,\\
		30 July- 6 August, 2015\\
		The Hague, The Netherlands}

\begin{document}

\section{Introduction}
We consider a quite cold old neutron star surrounded by a cloud of cold helium gas. Helium atoms in the gravitational field of the star are localized in long-lived quantum states, similar to the states of neutrons and antihydrogen atoms in the gravitational field of the Earth \cite{1,2}. Those states have already been studied theoretically. Experimental test of the existence of such states for antihydrogen by methods of induction of resonance transitions between quantum levels in temporally oscillating gradient magnetic field is planned \cite{3}. 

In case of dealing with helium atoms near the neutron star's surface neutron star's own oscillating magnetic field can be used to observe gravitational states of atoms. The main effect that makes these observations difficult is the thermal motion of helium atoms. Helium atoms were chosen because the distance between gravitational levels of helium is larger than the same for hydrogen, that's why the thermal motion's effects are not so drastic for helium. It is shown in the following paper that temperature about $0.4$ K will make observation of spectral lines consistent with gravitational transitions possible. Temperature about $0.4$ K could not be achieved if a case of weak anisotropy of cosmic microwave background is considered ($T_{\text{CMB}}\sim 2.7$ K)\cite{4}. On the other hand if we manage to register gravitational states' spectral lines, we can state the existence of the Universe's areas with sufficiently lower temperatures ($T\ll T_{\text{CMB}}$). 

The following work becomes the work of significant importance due to the preparing study of radiation with gigahertz frequencies \cite{5}. 
\section{Helium atom in a neutron star's field}
The behavior of helium atom near the neutron star's surface is considered. The surface is supposed to be an ideal one. Gravitational field of the neutron star is thought as linear with the potential of atom near the star's surface having the form $V(z)=M_{He}gz$, where $z$ is a height of the atom above the mirror, $g$ is a gravitational field intensity near the star's surface.

Energies $E_n$ and heights $Z_n$ of quantum gravitational states of helium atoms are (analogous to neutrons' states \cite{1})
\begin{equation}\label{Ener1}
E_{n}=\epsilon_{0} \lambda_{n},
\end{equation}
where $\epsilon_0=\sqrt[3]{\frac{M_{He}^2 g^2 \hbar^2} {2 m_{He}}}$
and $Ai(-\lambda_n)=0$; $Ai(x)$ - is an Airy function,
\begin{equation}\label{Heig1}
Z_n=l_0 \lambda_n,
\end{equation}
where $l_0=\sqrt[3]{\frac{\hbar^2}{2 M_{He} m_{He} g}}$, 
$M_{He}$ - gravitational mass of helium atom, $m_{He}$ - inertial mass of helium.

We consider the neutron star with the mass equal to $m_{ns}=1.4 m_{solar}$  and radius  $R_{ns}=6$ km ($R_g/R_{ns}=0.69$, $R_g$  - Schwarzschild's gravitational radius). For such a star the gravitational acceleration is
\begin{equation}\label{Grav}
g=\frac{G m_{ns}}{R_{ns}^2 \sqrt{1-R_g/R_{ns}}}=9.2614\cdot 10^{12}(\text{m/s}^2),
\end{equation}
$G$ - the gravitational constant.\\
Numerical values of gravitational energies and heights:
\begin{eqnarray}\label{CharNum}
\epsilon_0&=&91.920\mbox{ } \mu\mbox{eV},\\
l_0&=&2.3744\cdot 10^{-1} \mbox{nm}.
\end{eqnarray}
There are energies of the first five gravitational quantum states of helium atoms in Table \ref{Table 1}.

\begin{table}[hear]
\centering
\begin{tabular}{|p{5em}|p{7em}|p{7em}|}
\hline
$n$ & $\lambda_n$ & $E_n$, $\mu$eV \\
\hline
1 & 2.338 & 214.92 \\
2 & 4.088 & 375.76\\
3 & 5.521 & 507.45\\
4 & 6.787 & 623.83\\
5 & 7.944 & 730.22\\
\hline
\end{tabular}
\caption{The eigenvalues and gravitational energies of a helium atom in the neutron star's gravitational field.}
\label{Table 1}
\end{table}

Helium atoms undergo thermal motion, but we need this motion not to intermix different gravitational states. This requirement is quite essential to observation of transitions between quantum states in magnetic field of the star. 

It is rational to consider transitions between the first (ground) and the second gravitational quantum levels, because the distance between these levels is the largest. If the temperature is lower than $0.4$ K then the condition  $(E_2-E_1)>2E_T$  is satisfied ($2E_T/(E_2-E_1)=0.6$), where $E_T=3/2 kT$.

\begin{figure}
 \centering
\includegraphics[width=1\textwidth]{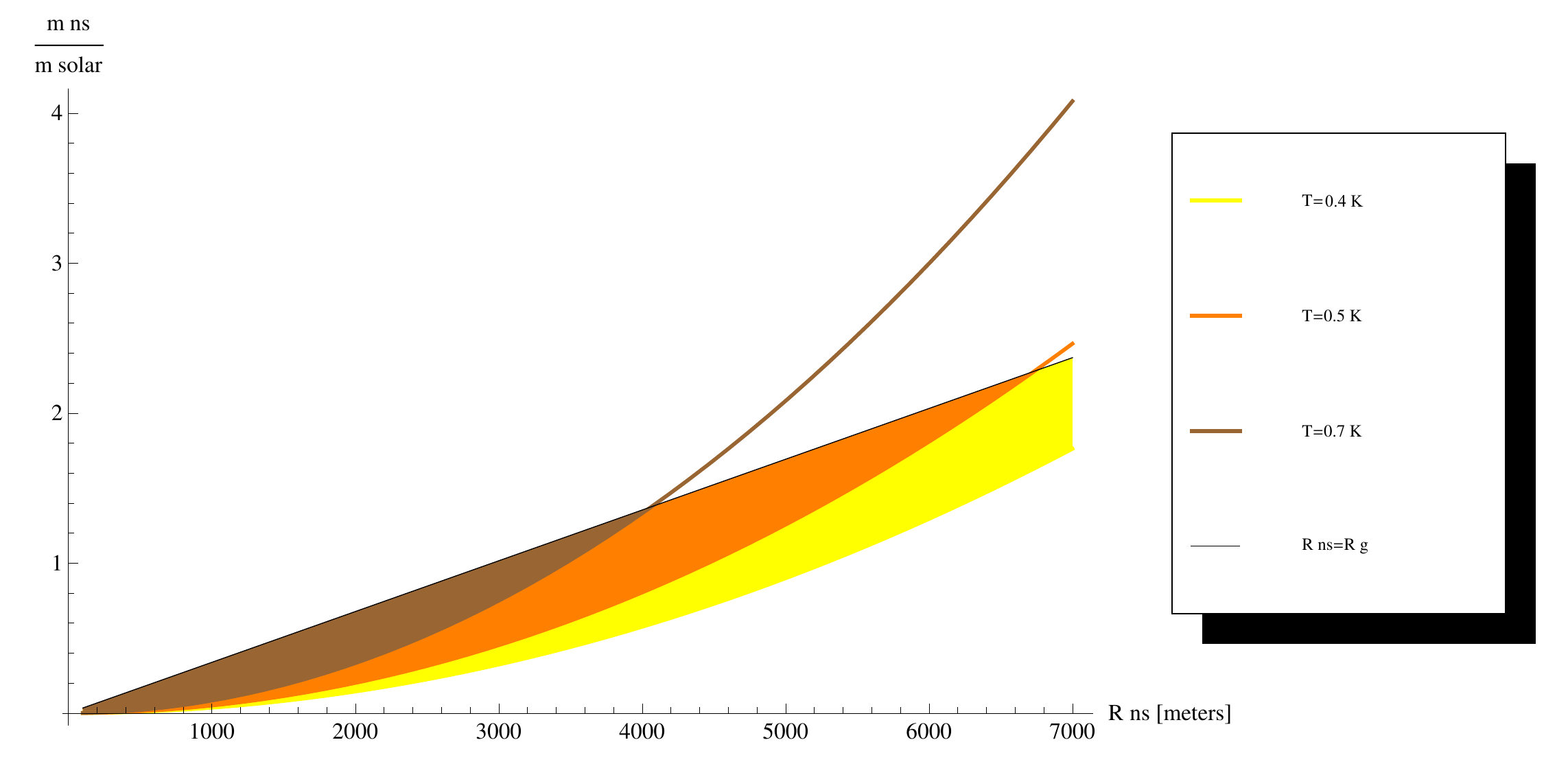}
\caption{Possible parameters (mass $m_{ns}$ and radius $R_{ns}$) of neutron star at different temperatures for which there is no thermal intermixing of the first and the second gravitational states of helium atom (filled area).}\label{FigParam}
\end{figure}

The essential requirement of the existence of the system neutron star - helium in the ground gravitational state is the condition of smallness of atom size (diameter of the second Bohr's orbit $2 R_B$) compared with the characteristic height of the ground state of atom $Z_1$  above the surface. The mentioned condition ($2R_B/Z_1=0.38$) is true.

Neutron star's magnetic field can be used to observe quantum gravitational states of helium atom. It will be shown that spatially inhomogeneous temporally oscillating magnetic field will induce transitions between gravitational states with nonzero probability. 

In some approximation the magnetic field of neutron star can be considered as the magnetic field of dipole:
\begin{equation}
\vec{B}=\frac{3\vec{n}(\vec{n},\vec{M})-\vec{M}}{r^3},
\end{equation}
where $M$ - dipole magnetic moment, $r$ - distance from the star's center, $\vec{n}$ - is a unit vector parallel to $\vec{r}$.

\begin{figure}
\centering
\includegraphics[width=.4\textwidth]{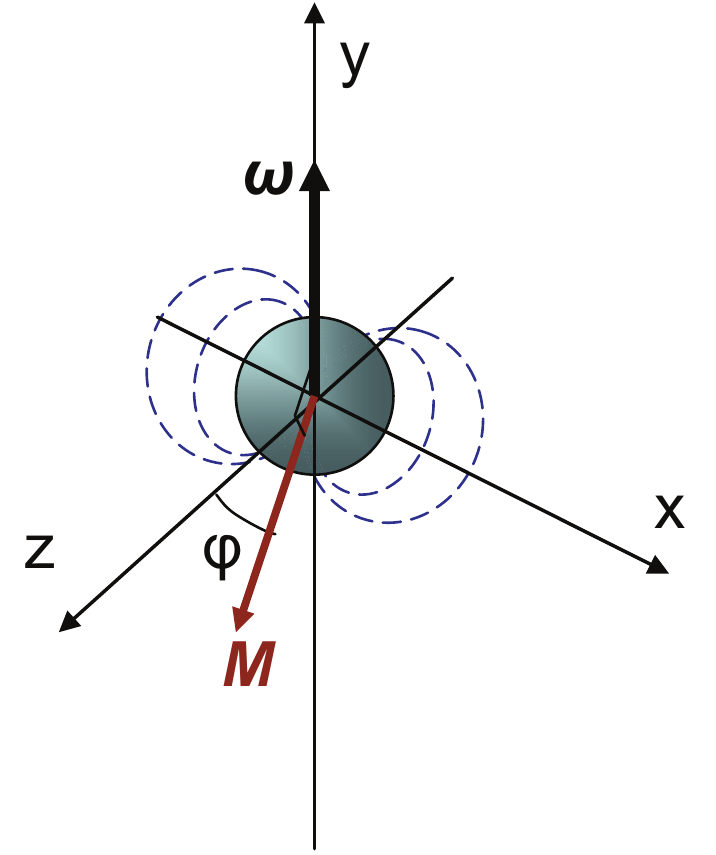}
\caption{Configuration of the magnetic field of the neutron star and the axis of its rotation in the chosen frame of reference.}\label{FigField}
\end{figure}

We consider magnetic field in a frame of reference shown in FIG.\ref{FigField}:
\begin{equation}
\vec{B}=\frac{2M\cos(wt)\vec{e}_z-M\sin(wt)\vec{e}_x}{r^3}.
\end{equation}
where $w$ - frequency of star's rotation, related to the period: $w=2\pi/T$.
\\
If we consider a normal neutron star, then it can be assumed:
\begin{eqnarray}
T&=&0.1 \mbox{ } \mbox{s},\\
B&=&10^{13}\mbox{G}.
\end{eqnarray}

Inhomogeneous magnetic field couples the spin and the spatial degrees of freedom of the atom. Influence of the magnetic field on the atom can be described by the operator: $\hat{V}_{field}=-\hat{\vec{\mu}}_{atom}\vec{B}$, where  $\mu_{atom}$ - magnetic moment of the atom. We assume helium to be in a  $2^3 S$-state: $1s^1 2s^1$. This state is characterized by a full spin $S=1$  and three possible spin projection quantum numbers: $S_z=-1,0,1$.

Hamiltonian $\hat{H}_0$ describing helium atom in the gravitational field of the star without magnetic field consists of two terms. One is the Hamiltonian of interatomic motion, $\hat{H}_{rel}$; another one is the Hamiltonian of center of mass motion, $\hat{H}_{cm}$. 
\begin{equation}
\left\{
\begin{array}{l}
\hat{H}_0=\hat{H}_{cm}+\hat{H}_{rel},\\
\hat{H}_{cm}=-\frac{\hbar^2}{2M_{He}}\frac{d^2}{dz^2}+M_{He}gz,\\
\hat{H}_{rel}=-\frac{\hbar^2}{2m_{e}}(\nabla_{\vec{R}_1})^2-\frac{\hbar^2}{2m_{e}}(\nabla_{\vec{R}_2})^2-\frac{Z e^2}{R_1}-\frac{Z e^2}{R_2}+\frac{e^2}{\left|\vec{R}_1-\vec{R}_2\right|}.
\end{array}
\right.
\end{equation}

Everywhere earlier $\vec{R}_i=\vec{r}_i-\vec{R}$,  $\vec{r}_i$ - coordinate of $i$-th electron $e_i$,  $\vec{R}$ - coordinate of nuclei, $z$ - coordinate of center of mass. 
 
Magnetic field's influence causes the appearance of a new term in the Hamiltonian:
\begin{equation}
\hat{H}=\hat{H}_0+2\mu_{B}\vec{B}(\hat{\vec{s}}_{e_1}+\hat{\vec{s}}_{e_2}).
\end{equation}

For shown configuration of magnetic field in the frame of reference connected with helium atom the operator of influence of magnetic field takes a form:
\begin{equation}
\hat{V}=\mu_{B}\frac{2M\cos(wt)(\sigma_{z1}+\sigma_{z2})-M\sin(wt)(\sigma_{x1}+\sigma_{x2})}{(z+R_{ns})^3},
\end{equation}
where $\sigma_{i}$ are the Pauli matrices, $z$ is atom's height above the neutron star's surface.

We suppose that the magnetic moment is fixed along $z$-axis, i.e. atom's state is described by a spin projection quantum number $S_z=1$, that does not change.

The estimation of the possibility of transition between the first and the second gravitational states induced by the star's magnetic field (transition frequency $\nu_{12}=w_{12}/2\pi=3.8892\cdot 10^{10}$ Hz) is made.
 
The corresponding Schr\"{o}dinger equation has a form:
\begin{equation}
-\frac{\hbar}{i}\frac{d}{dt}\Psi=(\hat{H}_0+\hat{V})\Psi.
\end{equation}
The solution of the following equation is found in the form (using the two-state system model):
\begin{equation}
\Psi=C_1(t) e^{-i w_1 t}|1>+C_2(t) e^{-i w_2 t}|2>,
\end{equation}
where $|1>$, $|2>$ - eigenvectors of the Hamiltonian $\hat{H}_0$  corresponding to the gravitational states 1 and 2 ($|n>=Ai(z/l_0-\lambda_n)/N_n\cdot  \psi(R_1,R_2)\chi^{1}$, where $N_n$ is a normalizing coefficient for the Airy function $Ai(z/l_0-\lambda_n)$, $\psi(R_1,R_2)$ is a space-dependent wave function of helium atom, $\chi$ is a spin wave function).

The initial condition for our system is the following: at the initial moment of time atom is in the ground gravitational state, $C_1(0)=1$, $C_2(0)=0$.

\begin{figure}
 \centering
\includegraphics[width=.8\textwidth]{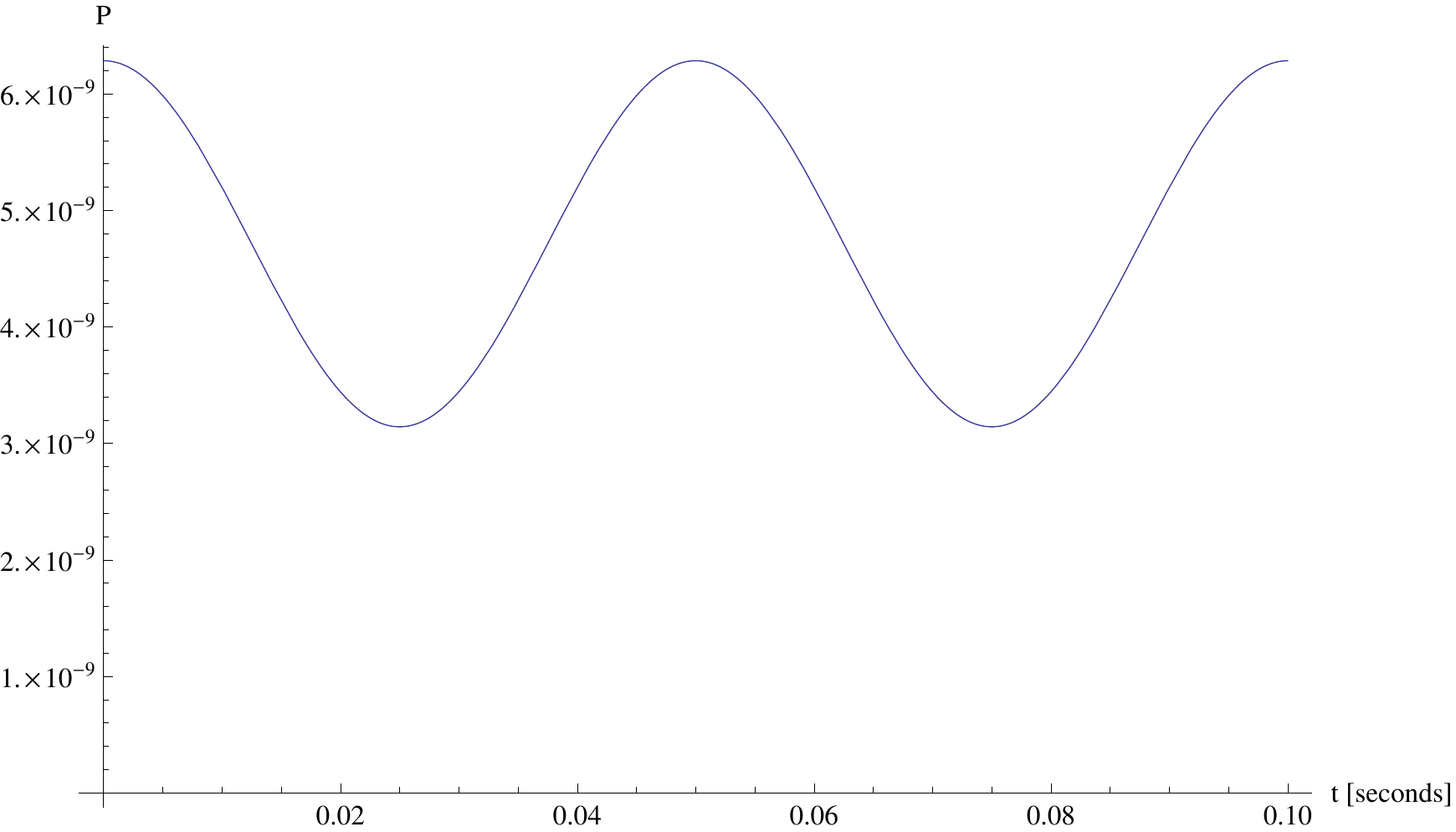}
\caption{Average value of the transition probability as a function of time.}\label{Probab2}
\end{figure}

One can try to find the possibility of transition between gravitational quantum states using the formalism of perturbation theory. In this case from the initial conditions one can find the "zero approximation" for the coefficients:
\begin{equation}
\left\{
\begin{array}{l}
C_1^{(0)}=1,\\
C_2^{(0)}=0.
\end{array}
\right.
\end{equation}
The full solution looks like the following:
\begin{equation}
C_n=C_n^{(0)}+C_n^{(1)}+C_n^{(2)}+\dots.
\end{equation}
Keeping only the first two terms and using the condition $w\ll w_{12}$, one can get the probability of transition between the first and the second gravitational states  $|C_2|^2$:
\begin{equation}
P=\frac{W_{12}^2}{(\hbar w_{12})^2}(1-2\cos(wt)\cos(w_{12}t)+\cos^2(wt)).
\end{equation}
After averaging over time interval  $\tau=\pi/w_{12}$, one will get (see FIG.\ref{Probab2}):
\begin{equation}
P=\frac{W_{12}^2}{(\hbar w_{12})^2}(1+\cos^2(wt)),
\end{equation}
where matrix elements have the form: $W_{nk}=<n|4\mu_{B}\frac{M}{(z+R_{ns})^3}|k>$.

Occurred transitions can be registered by analyzing star's spectrum (one can detect the line in spectrum consistent with the transition frequency). The frequency of photons emitted during the transition near the surface of the star in a locally inertial reference frame $\nu_{12}=3.8892\cdot 10^{10}$ Hz  will undergo gravitational redshift when leaving the star and become  $\nu_{\infty}$ :
\begin{eqnarray}
z_g&=&(1-R_g/R_{ns})^{-1/2}-1,\\
\nu_{\infty}&=&\nu_{12}/(1+z_g)=2.1679\cdot 10^{10} \mbox{Hz}.
\end{eqnarray}

\section{Intensity of gravitational transition's spectral line}
One needs to compare the intensity of gravitational transition's spectral line with the intensity of thermal emission to be sure that the first one is observable. Spectral line arises from the spontaneous emission of helium atom in the excited second gravitational state. 

According to the perturbation theory the probability of spontaneous emission in a unit of time is determined as     

\begin{equation}
dP_{fi}=\frac{2\pi}{\hbar}|<f|\hat{U}|i>|^2 d\rho(E_f),
\end{equation}
where $|i>$, $|f>$ are the initial and the final states of the system atom-electromagnetic field respectively, $d\rho(E_f)$ is the density of the number of final states, $\hat{U}$ is the interaction operator of the field (characterized by the magnetic vector potential $\vec{A}$) with the magnetic moment of the atom having the form:
\begin{equation}
\hat{U}=-\vec{\mu}_{atom}rot\vec{A}=2\mu_{B}(\vec{s}_{e_1}+\vec{s}_{e_2})rot\vec{A}.
\end{equation}
In a case of emission of one photon the interaction operator is 
\begin{equation}
\hat{U}=\sum_{\alpha}2\mu_{B}(\frac{2\pi\hbar c^2}{V w})^{1/2}e^{-i\vec{k}\vec{r}-iwt}i(\vec{s}_{e_1}+\vec{s}_{e_2},[\vec{k},\vec{e}_{\alpha}])a^{+}_{\vec{k},\alpha},
\end{equation}
where $\vec{k}$ is a wave vector of emitted photon, $k=w/c$, $\vec{e}_{\alpha}$ is its polarization vector, $a^{+}_{\vec{k},\alpha}$ is the photon creation operator.

The initial and the final states of the system atom-electromagnetic field are 
\begin{equation}
|i>=|n_{\vec{k},\alpha}>{Ai(z/l_0-\lambda_2)}/{N_2} \cdot\chi^{1}=|0>|2> \chi^{1},
\end{equation}
 \begin{equation}
|f>=|1>|1> \chi^{1},
\end{equation}
as we assumed helium to be in a  $2^3 S$-state: $1s^1 2s^1$ with a spin projection quantum number $S_z=1$, that did not change; $|n_{\vec{k},\alpha}>$ is a state in which $n$ photons are in the mode $\vec{k},\alpha$.

After the summation over $\alpha$ and integration over solid angle one can get the probability of spontaneous emission in a unit of time
\begin{equation}
P_{21}=\frac{16 \mu_{B}^2 w_{21} g^2}{15 c^5 \hbar}.
\end{equation}
The intensity of the spectral line 
\begin{equation}
I_{21}=\hbar w_{21} P_{21}\sim 10^{-28} \text{eV/s}.
\end{equation} 
Atoms of helium in the second gravitational state form a monatomic layer above the surface of the star, so the number of them
\begin{equation}
N_{He}=\frac{S_{ns}}{S_{He}}=\frac{4\pi R_{ns}^2}{\pi r_{He}^2}\sim 10^{28},
\end{equation}  
and the total intensity of them
\begin{equation}
I_{N}=N_{He}I_{21}\sim 1 \mbox{ } \text{eV/s}.
\end{equation}

The lifetime of helium atom above the surface of the star in the second gravitational state is determined by the process of inelastic scattering of helium from neutrons of the surface. The following lifetime $\tau_2$ is
\begin{equation}
\frac{1}{\tau_2}=v n_{xy}|\psi(z=r_{nucl})_2|^2\sigma_{inelast},
\end{equation}  
where $v=\sqrt{2 E_2/m_{He}}$ is the speed of helium atom falling to the surface, $n_{xy}$ is the density of helium atoms in $xy$ plane, $\psi(z=r_{nucl})_2$ is the wave function of helium at the nuclear distance $r_{nucl}=2\cdot 10^{-13}$ sm, $\sigma_{inelast}$ is the inelastic scattering cross section of neutron from alpha-particle at low energies. The cross section is known to be less than $\sigma_{inelast}\lesssim 10^{-4}$ b according to the Nuclear Data library RUSFOND.

The lifetime $\tau_{2}$ and the corresponding spectral line width $\delta \nu$ are 
\begin{eqnarray}
\tau_{2}&\gtrsim&8.7831\cdot 10^{10}\mbox{s},\\
\delta \nu&\lesssim& 1.1386\cdot 10^{-11} \mbox{Hz}.
\end{eqnarray}

The thermal emission of the neutron star considered as the black body with the temperature $T=0.4$ K has the intensity according to the Planck's law:
\begin{equation}
I_{ns}=\frac{\hbar w_{21}^3}{4\pi^2 c^2}\frac{1}{e^{\hbar w_{21}/kT}-1} S_{ns}\delta \nu\lesssim 10^{-4} \text{eV/s},
\end{equation}  
where $S_{ns}$ - the surface area of the neutron star, $\delta \nu$ - width of the spectral line due to the interaction with the surface.

If we consider the detector on the Earth with linear scale $l_{det}\sim 1$ km and assume that the distance between the star and the detector is $L=5000$ light years (the distance between the Earth and the Boomerang nebula, which is the coolest place known in the Universe), then the intensity of radiation to the given solid angle will be:
\begin{equation}
I_{N, d\Omega}=\frac{1}{4\pi}I_{N}\frac{\pi l_{det}^2}{L^2}\sim 10^{-34} \text{eV/s}.
\end{equation}   

The intensity of cosmic microwave background radiation ($T_{\text{CMB}}=2.7$ K) to the detector:
\begin{equation}
I_{CMB}=\frac{\hbar w_{21}^3}{4\pi^2 c^2}\frac{1}{e^{\hbar w_{21}/kT_{\text{CMB}}}-1} (\pi l_{det}^2)\delta \nu\lesssim 10^{-4} \text{eV/s}.
\end{equation}  

\section{Conclusion}

Gigahertz range of radiation from neutron stars has not been investigated well till now, but in the near future such observations are planned. Registration of spectral lines consistent with transitions between gravitational levels is of great interest. If gravitational states' spectral lines are detected, we will be able to study quantum gravitational states of helium, to get gravitational mass of the atom and to get new information about the temperature of cosmic microwave background and the Universe. 

\acknowledgments

The authors would like to thank A.Yu. Voronin and V.V. Nesvizhevsky for very useful discussions.

\end{document}